# Deeply fused flow and topology features for botnet detection based on a pretrained GCN

Xiaoyuan Meng[a,1*], Bo Lang[a,b,1*], Yanxi Liu[a,c], Yuhao Yan[a]

[a] State Key Laboratory of Software Development Environment, Beihang University, XueYuan Road, Beijing, 100191, Beijing, P. R. China.

[b] Zhongguancun Laboratory, Beijing, P. R. China.

[c] China Mobile Information Technology Co., Ltd.

[1] These authors contributed equally to this work.

*Corresponding authors. E-mail(s): mengxiaoyuan@buaa.edu.cn (Xiaoyuan Meng), langbo@buaa.edu.cn (Bo Lang).

## Abstract

The characteristics of botnets are mainly reflected in their network behaviors and the intercommunication relationships among their bots. The existing botnet detection methods typically use only one kind of feature, i.e., flow features or topological features; each feature type overlooks the other type of features and affects the resulting model performance. In this paper, for the first time, we propose a botnet detection model that uses a graph convolutional network (GCN) to deeply fuse flow features and topological features. We construct communication graphs from network traffic and represent node attributes with flow features. The extreme sample imbalance phenomenon exhibited by the existing public traffic datasets makes training a GCN model impractical. To address this problem, we propose a pretrained GCN framework that utilizes a public balanced artificial communication graph dataset to pretrain the GCN model, and the feature output obtained from the last hidden layer of the GCN model containing the flow and topology information is input into the Extra Tree classification model. Furthermore, our model can effectively detect command-and-control (C2) and peer-to-peer (P2P) botnets by simply adjusting the number of layers in the GCN. The experimental results obtained on public datasets demonstrate that our approach outperforms the current state-of-the-art botnet detection models. In addition, our model also performs well in real-world botnet detection scenarios.

# 1 Introduction

A botnet consists of multiple infected nodes and one or more master nodes. Under the control of the master nodes, compromised nodes execute various malicious behaviors. However, simultaneously, these nodes maintain their regular functions, which makes botnets cyberattacks that are difficult to detect. The carriers of botnets are diverse and include personal computers, Android systems, and the Internet of Things (IOT) [1][2]. It is evident that botnets steal users' information, but the harm extends beyond this. Compromised nodes are also exploited to distribute malicious software, send spam emails, and even initiate distributed denial-of-service (DDoS) attacks[3][4][5][6]. The master nodes communicate with the infected nodes via a command-and-control channel (C2C). Based on their C2C architectures, botnets are classified as either C2 or peer-to-peer (P2P) networks. In the former, nodes are explicitly categorized as master nodes or controlled nodes. However, in P2P networks, compromised nodes may also have the capacity to control other nodes [7][8].

As botnets and legitimate nodes are behaving increasingly alike, it becomes challenging to detect botnets in real time with high precision. The current detection methods can be broadly classified into flow-based and graph-based methods, each of which focuses on different aspects, i.e., network behavior characteristics and the topology attributes of communication graphs, respectively.

Flow-based studies center on network behaviors. Network flow features, such as communication frequencies and packet lengths, are analyzed to distinguish bot nodes [7][8][11][19][20][26]. New types of botnets often employ the domain generation algorithm (DGA) or fast-flux (FF) technique to conceal themselves; thus, many studies have focused only on domain name resolution traffic [9][10]. These studies deploy machine learning models or neural networks for node classification purposes. However, these methods ignore the topological characteristics that provide insights into the broader structures of networks.

In contrast, studies based on network communication graphs [12][13][21] utilize feature engineering to extract topological features such as in-degree, out-degree, centrality, and PageRank features. Afterward, supervised or unsupervised machine learning models are applied to detect botnets. However, the process of extracting topological features is time-consuming, especially when the network scale is expanded. Therefore, topology-based methods using feature engineering are unsuitable for real-time detection.

Recent advancements in graph neural networks (GNNs)[24] have promoted research on GNN-based botnet detection approaches. These kinds of methods [17][23][29] apply a graph convolutional network (GCN) or graph isomorphism network (GIN) to learn the deep topological representation of the target communication graph and utilize a fully connected layer to classify the graph nodes. However, in real networks, there are far more legitimate nodes than malicious nodes, causing severe imbalances in the publicly available botnet datasets. For instance, in the CTU-13 dataset [24], the ratio of malicious to benign samples is even 1:25000. Consequently, GCN models trained on such datasets tend to exhibit poor recall rates, which means that these models are invalid. Therefore, model training becomes a main obstacle preventing the application of GCN models. Recently, GNN-based methods [26][36] have been proposed for training GCNs and classifying them on artificially balanced graph datasets [17]. However, the characteristics of artificial datasets usually cannot be guaranteed to be consistent with real traffic, which affects the performance of these models to a great extent when running in real circumstances.

Flow features and topological features separately offer unique insights into botnets.

However, relying exclusively on either flow or topological features may lead to the loss of additional valuable information. The combination of these two types of features enhances the effect of the constructed detection model. When a GCN model processes graph data, it utilizes both node features and internode connection relationships [15]. In light of this, we propose a real-time botnet detection method based on a GCN model to deeply fuse flow features and topological features. For the flow features, we choose five features that are easily obtained and are not specific to a certain type of botnet. These features encapsulate the behaviors of the network nodes and are represented as the node features of the communication graph. For the topological features, we employ the GCN to mine the characteristics of the communication graph topology. The GCN also fuses the flow features and topological features.

To mitigate the limitations of training a GCN on imbalanced datasets and avoid the problem of excessively training the model by relying on artificial datasets, we implement an efficient pretraining strategy. Our detection framework consists of a pretrained GCN model and a classification model. Initially, we train the GCN on a balanced graph dataset as applied in [17], which allows the GCN to extract topological features; then, we train a classification model that takes the features derived from the GCN as inputs by using the data from the original datasets. In the detection phase, the communication graph with flow features is fed to the pretrained GCN. The output of the final hidden layer is taken as the fused features and is subsequently input into an Extra Tree model for node classification purposes. Therefore, compared with other GCN-based models, our proposed method renders a comprehensive, multifaceted representation of the communication characteristics of the nodes, and its performance can be further improved by training the Extra Tree model on the original datasets. Additionally, to adapt to the variations exhibited by botnets under C2 and P2P architectures, our model can maintain satisfactory detection effects for these two types of botnets by adjusting the number of utilized GCN layers.

The main contributions of this paper are as follows.

(1) We effectively fuse the flow features and topological features of nodes by using a GCN and propose rules for flow feature selection. Based on traffic, we construct communication graphs that comprehensively represent network behaviors and the internode communication relationships and utilize the information aggregation ability of a GCN to achieve deep feature fusion. We also propose rules for flow feature selection and choose five effective flow features.

(2) To make the GCN training process effective, we propose a strategy of pretraining the GCN to establish a new detection framework. To achieve valid GCN training on extremely imbalanced flow datasets, we first propose pretraining the GCN rather than jointly training the GCN and the classification model on the artificial balanced graph datasets. Additionally, our GCN model can be easily applied to detect botnets with C2 and P2P architectures by adjusting only the number of GCN layers.

(3) The experimental results show that our approach outperforms the current state-of-the-art botnet detection models and has good performance in real-world botnet detection scenarios.

The remainder of this paper is organized as follows. Section 2 introduces the related work. Section 3 introduces our method. Section 4 presents the results of our experiments. Section 5 summarizes our work.

## 2 Related work

Due to the unique behavior patterns of bots, botnet detection is commonly categorized as an anomaly detection problem [18]. According to the type of data targeted for detection, botnet detection methods are generally classified into two categories: flow-based and graph-based detection methods.

Flow-based approaches mainly analyze the characteristics of node traffic. These methods slice the original traffic flow over time and extract features within a fixed time window [26]. Consequently, flow-based methods inherently possess real-time detection capabilities. Kirubavathi et al. [11] targeted small packets by extracting four features and achieved commendable results with a naive Bayesian classifier. Since new botnets often employ the HTTP protocol and DGA technique, Udiyono et al. [34] utilized features such as the number of HTTP POST requests and the frequency of DNS server queries to classify botnets using a random forest model. With the rapid development of neural networks, a growing number of methods have begun utilizing neural networks for classification purposes, but their central focus is still on feature engineering. van Roosmalen et al. [33] focused solely on TCP/UDP packets, utilizing 13 packet-based features, including ACK flag and time-to-live (TTL) features. Afterward, they trained a deep neural network for classification. Alauthaman et al. [19] extracted 29 connection-based features from the headers of TCP control packets, including the average duration of a single connection and the average numbers of packets and bytes transferred in each connection. A decision tree was used for feature selection, and a neural network was trained for classification. However, some scholars have noted [21] that flow-based detection often assumes that the network behaviors of bots and legitimate nodes are significantly different. However, as botnets evolve, they behave more like normal nodes, and their distinctness, which is reflected in flow features, gradually decreases.

Graph-based methods focus on the global structures of botnets by analyzing the topological features embedded within the target network communication graph. Researchers such as Abou et al. [12] and Alharbi et al. [21] extracted the in-degree, out-degree and centrality of nodes as features and then employed machine learning models or neural networks for classification. Chowdhury et al. [13] used topological features, including node centrality and clustering coefficients, to lock the bot nodes into a smaller cluster through clustering. However, such methods require network traffic to be collected over a longer period, and the time cost of feature extraction can increase substantially as the network scale increases. It has been noted in [22] that traditional topological features such as in-degree and out-degree features are coarse-grained features, and it is difficult to capture the fine-grained interaction behaviors between nodes. In recent years, graph neural networks (GNNs) have been applied to botnet detection. Zhou et al. [17] used the message propagation mechanism of GCN models to learn a deep representation of the topological features of a network. By constructing communication graphs and utilizing a 1st-order unit matrix as the node features, they trained a GCN to perform node classification. Lo et al. [16] devised a GIN with grouped reversible residual connections, which further improved the detection effect of their model. Nonetheless, the effect of a GNN relies strongly on the degree of balance exhibited by the utilized training dataset. Both Lo et al. [16] and Zhou et al. [17] trained GNNs with artificially balanced graph datasets. Currently, methods based on GNNs do not use traffic features.

Recently, some studies have recognized the significance of combining topological and flow features. Wang et al. [23] designed three models: two models based on flow features and another model based on topological features. The nodes and classified nodes were scored by each model using a preset threshold. Finally, the bots were determined by voting on these three classifiers, which resulted in better performance than that achieved by using either topological features or flow features alone. However, this approach does not deeply fuse these two types of features, and the model requires prior knowledge of the number of bots contained in the network. Hence, its detection effect degrades when the network environment is completely unknown.

# 3 Method

## 3.1 Motivation and framework

In botnets, each node has specific flow features. Simultaneously, node communication relationships possess topological features. Flow-based detection methods do not require long-term network traffic. They only need the network flow information contained within a fixed time window, effectively supporting real-time detection [26]. Furthermore, the use of a sliding time window can minimize the possibility of a single malicious behavior instance being split into two windows, which would affect the detection results. Although flow feature utilization has efficiency advantages, designing effective flow features is more difficult when botnets are increasingly covert. Conversely, topology-based methods can accurately identify bots but require long-term network information and a greater amount of time [21].

Considering the pros and cons of these two types of features, we propose fusing them to integrate their advantages and establish a new detection model.

We construct network communication graphs by storing flow features as node features, thereby simultaneously representing the connection topology and the node behavior information of the input graph data. In recent years, GNNs have become powerful tools for processing graph data. GCN models can pass node features to the surrounding nodes through a message propagation mechanism. This enables GCNs to learn the topological information of the graph and simultaneously fuse the node features with the topological features. Additionally, nodes connected to the bot nodes are more likely to be bots, and the message propagation mechanism of a GCN enables the nodes to 'sense' the flow behaviors of their neighboring nodes. Hence, each node can obtain deep features that aggregate the behaviors of other nodes. Therefore, we decide to use a GCN to extract topological features and fuse them with the flow features of the graph nodes.

The overall framework of our method, as shown in Figure 1, is composed of four parts: a time slice and flow filter, a flow feature extraction module, a topological and flow feature fusion mechanism, and a node classification module.

The first stage involves time slice and flow filtering. Owing to the growing volume of network traffic [19], a suitable flow filtering technique is necessary. Currently, the TCP and UDP protocols are utilized predominantly in botnets [4]. Hence, retaining the traffic related to these protocols is a simple yet effective filtering strategy. Then, real-time traffic is split into overlapping slices using a sliding time window, and only the TCP- and UDP-related traffic within each window is retained. Notably, the sizes and intervals of the sliding time windows markedly influence the detection effect of the model. Motivated by Alauthman et al. [7], we adopt the setting of a *60 s* window size and *10* s intervals to ensure that detection can be implemented in real time while achieving good detection performance.

The second stage is flow feature extraction. We extract five essential flow features from the filtered network traffic, which include the number of transmitted bytes and the number of established connections. These flow features reflect basic network behaviors.

To balance the detection effect of the model with efficiency, we use features that do not necessitate complex computations. A more detailed analysis is presented in Section 3.2.

The third stage concerns the fusion of topological and flow features. To extract topological features, we construct a communication graph with the flow features derived from the input network traffic. We use a balanced graph dataset [17] to pretrain a GCN for topological feature extraction purposes. A communication graph is fed into the pretrained GCN, whose parameters are frozen. The output of the final hidden layer serves as the fused features. The detailed descriptions of these steps are elucidated in Section 3.3.

The fourth stage is node classification. We consider a range of widely used and outstanding machine learning models, including ensemble learning models, support vector machines (SVMs), and multilayer perceptrons (MLPs). We finally choose an ensemble learning model, Extra Tree [30], which offers the best performance.

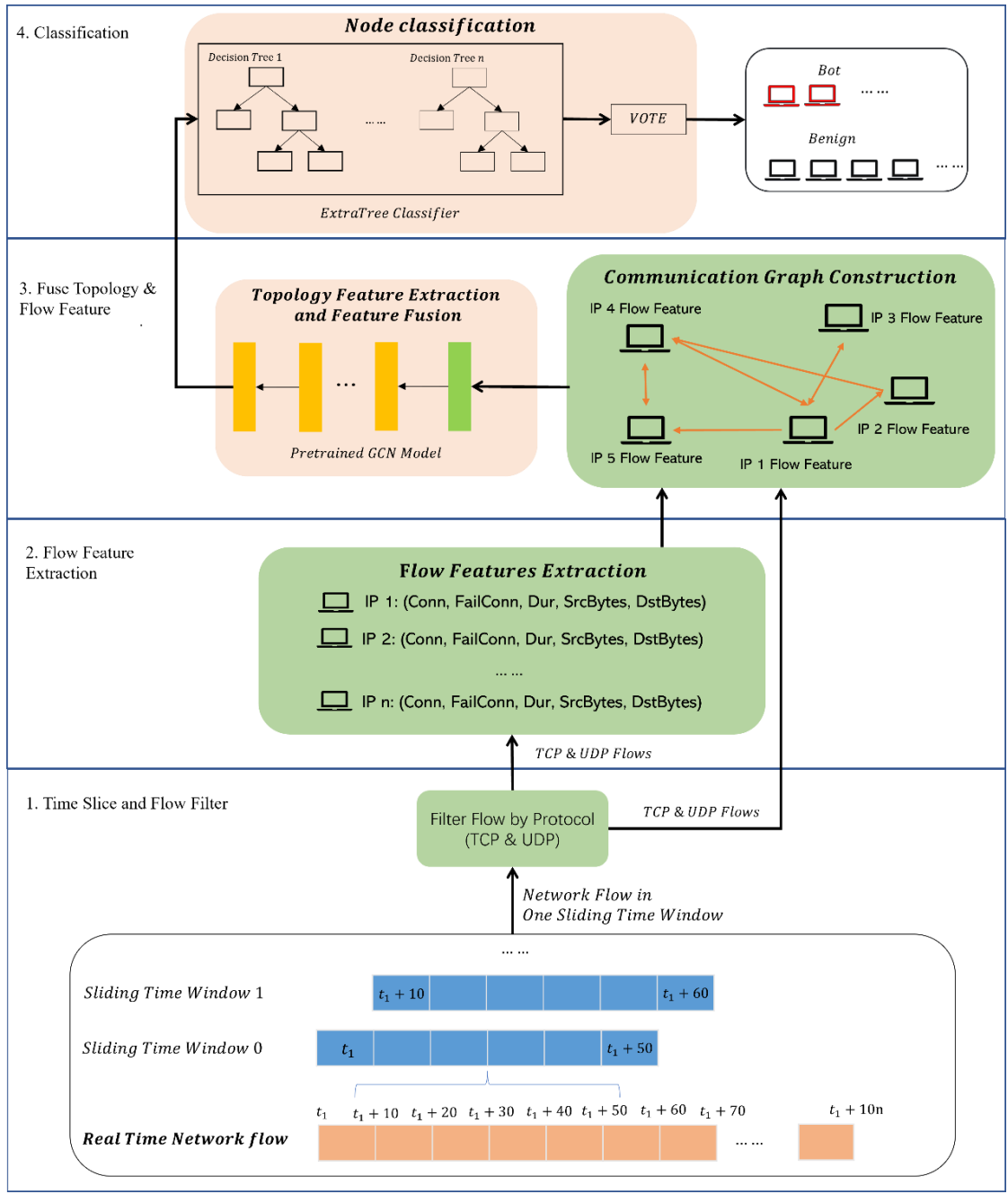

Figure 1 The framework of the proposed detection model.

## 3.2 Flow feature extraction

We identify a five-tuple (SrcIP, DstIP, SrcPort, DstPort, Proto) as a communication flow, where SrcIP and DstIP represent the source IP and destination IP, respectively, SrcPort and DstPort represent the source port and destination port, and Proto indicates the protocol used in this communication process. We extract flow features for each node.

In the literature, several fundamental flow features have been widely used. These include the number of transmitted packets, the packet length, and the number of connections. Several researchers [19][20][26] have adopted both elementary features and their statistics, for instance, the variance of the transmitted packets and bytes or the ratio between the numbers of sent and received packets. However, this approach requires additional computational resources and must constantly track traffic packet sequences. Through experiments, we find that without incorporating these statistical features, satisfactory detection performance can also be obtained. [11] concentrated on only four features related to small packets. Although the features of small packets are representative, they are easily evaded by attackers (such as by adding irrelevant payloads and extending the lengths of packets).

Taking this analysis into account, we determine two rules when selecting features.
(1) To minimize both the time and computational costs, the utilized features should be easily accessible and require as few extra computations as possible.
(2) We avoid using features that are easy to evade, including features for small packets or features related to specific packet fields.

Based on these two rules, we select the following 5 features.

- **Total number of successful connections (Conn):** This feature describes the total number of successful connections established by a node. The infected nodes in a botnet need to send heartbeat packets regularly to inform the master of their presence, which means that infected nodes often have more connections.

- **Total number of failed connections (FailConn):** This feature describes the number of failed connections possessed by a node. A typical malicious behavior exhibited by a botnet node involves collecting vulnerable port information through port scanning, which generates far more failed connections than legitimate nodes.

- **Average communication flow duration (Dur):** This feature reflects the average duration of the successfully established connections. Most of the malicious connections are used only for network sniffing functions whose durations is very short. Consequently, the average connection durations of the bots are shorter than those of legitimate nodes.

- **Average number of sent bytes (SrcBytes) and average number of received**

bytes (DstBytes): These two features reflect the average numbers of bytes sent and received by each node. The sizes of malicious packets, such as heartbeat packets or port scanning packets, are mostly small. The volume of data contained in malicious packets is often lower than that in legitimate packets (such as file downloads and video transmissions). Thus, these two features are important for distinguishing bots from normal nodes.

## 3.3 Topological feature extraction and feature fusion based on a GCN

### 3.3.1 Model design

We construct communication graphs based on the intercommunication relationships of the nodes in the network. The flow features depicted in Section 3.2 are taken as the attributes of the nodes in the communication graphs. Subsequently, we employ the GCN for topological feature extraction and feature fusion.

A communication graph can be defined as a directed unweighted graph $G = (V, E)$, where $V$ represents the set of nodes in the network. Each node is identified by its IP address and is denoted as $V_{IP}$. We use $F_{IP}$ to represent the flow features of each node, where $F_{IP} = (Conn, FailConn, Dur, SrcBytes, DstBytes)$. $E$ is a set of directed edges representing the communication relationships among the nodes. For each connection, when $SrcBytes \neq 0$, $V_{SrcIP} \rightarrow V_{DstIP}$ is added to $E$; when $DstBytes \neq 0$, $V_{DstIP} \rightarrow V_{SrctIP}$ is added to $E$.

The overall structure of the GCN model is shown in Figure 2. It primarily comprises an input layer, $N$ hidden GCN layers, and several residual layers. The dimensionality of the node features is 5, and this value is coordinated with the dimensionality of the flow features. The number of GCN layers, $N$, is contingent on the botnet architecture, which is elaborated upon in Section 3.3.3. In each residual layer, the inactivated features are merged with the activated features by addition[31]. The implementation of the residual layer enhances the classification accuracy of the model and mitigates the vanishing gradient problem.

Each GCN layer in the model uses a message propagation mechanism, aggregating the information of the adjacent nodes with an adjacency matrix. This mechanism enables the recognition of topological characteristics [14]. The message propagation function $prog(A)$ used in the GCN layers is expressed as Equation (1). Here, $A$ represents the adjacency matrix of the graph. $D$ is a diagonal matrix, where an element $d_{ii}$ on the diagonal line is the degree of node $i$.

$$prog(A) = D^{\frac{-1}{2}} A D^{\frac{-1}{2}} \qquad (1)$$

By defining the node features with the flow features, the GCN layers merge these flow features into the computation process. Consequently, we acquire a merged representation of the topological and flow features.

Given that the training procedure of the GCN heavily relies on dataset balance, we opt to use a manually processed balanced graph dataset [17] for GCN pretraining. This allows the GCN model to learn the topological botnet features. After performing pretraining, we freeze the weight parameters of each GCN layer. When a communication graph with flow features is fed into the GCN, the outputs of the final hidden layer are taken as the fused features for the nodes. In the pretraining stage of the GCN model, a fully connected layer is used for node classification. In contrast, when the model is employed for feature fusion and botnet detection, an Extra Tree classifier is attached for node classification.

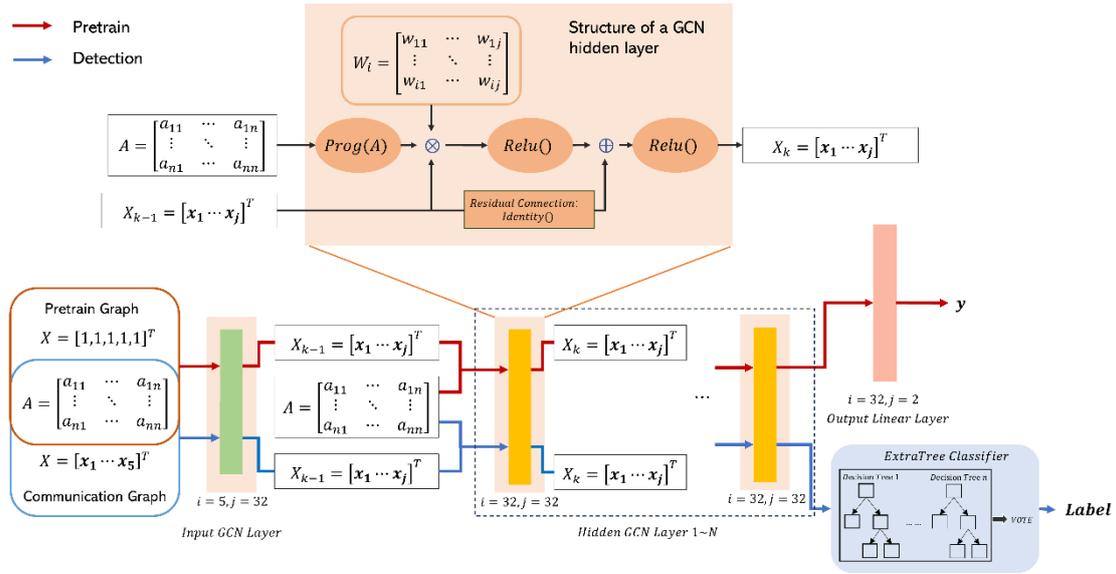

Figure 2 The structure of the pretrained GCN. Here, $A$ represents the adjacency matrix of the communication graph. $X$ denotes the feature vector of the nodes in the communication graph. $n$ represents the total number of nodes. $X_k$ represents the node embedding in the $k$th layer, $Prog(A)$ refers to the message propagation function introduced in Section 3.3.1, and $Relu()$ is the activation function. $i$ and $j$ represent the input and output dimensions of a layer, respectively.

### 3.3.2 Pretraining process of the GCN

During the pretraining phase of the GCN, we utilize the public graph dataset provided in [17]. This dataset is divided into two subdatasets: a botnet with a P2P architecture and a botnet with a C2 architecture. The authors removed all traffic flow information from the dataset, retaining only the graph connection information. The dataset has balanced numbers of positive and negative samples, making it suitable for GCN training.

As shown by the red box and arrows in Figure 2, during the pretraining phase, the node features form a vector of all ones, i.e., $x = [1,1,...1]$, meaning that the flow features are not introduced. In this case, changes in the weight matrix are induced only by the adjacency matrix. Thus, the GCN learns only the topological botnet features. During the training process, we use the early stopping strategy to prevent overfitting. The accuracy of the model during pretraining exceeds 98.9%, indicating that the model effectively learns topological features. After pretraining, the parameters of the GCN are frozen and do not change when running our botnet detection framework.

**3.3.3 Utilizing the GCN model for the detection of C2 and P2P botnets**

In the design of the GCN model, the number of layers significantly influences the model capacity. We hope that during pretraining, the GCN sufficiently learns topological botnet features. Moreover, the pretrained GCN can efficiently integrate flow features and topological features. A study in [14] demonstrated that for a simple topological feature learning task, merely adding 1-4 GCN layers can decrease the test loss to $10^{-3}$ of its initial value. This fact underlines that an increase in the number of layers can enhance the topological feature extraction capacity of the model.

The C2 and P2P botnet architectures are different. The former incorporates a central master, branching out in a star-like pattern. Conversely, the latter exhibits a mesh structure where each node potentially expands outward with infected nodes, making the structure more complex. Based on the structural characteristics of the two kinds of botnets, we believe that P2P botnets require more GCN layers than C2 botnets for extracting topological features using GCNs.

The message propagation mechanism allows the behaviors of malicious nodes to be transmitted to their surrounding nodes. With $k$ GCN layers, the messages from all nodes within $k$ hops are aggregated. The use of too many layers causes nodes that are far from the anomaly nodes to be affected during feature fusion. Considering this fact, the number of GCN layers must be controlled within a certain range.

Considering all the above considerations and the experimental results, we select a 12-layer GCN to extract the C2 botnet features and a 24-layer GCN to extract the P2P botnet features.

**3.3.4 Normalization of the fused features**

In high-speed networks, the traffic flow volumes are generally greater than those observed in low-speed networks. Hence, the statistical flow features of these two kinds of networks may have different distributions. Models trained with features derived from high-speed networks may not work well with low-speed networks. To prevent model failures caused by changes in the network environment, we perform min–max normalization on the fused features. This process converts each element of the feature vector to the 0-100 range. The conversion formula is shown as Equation (2):

$$\dot{x}_i = \frac{x_i - x_{min}}{x_{max} - x_{min}} \times 100 \qquad (2)$$

where $\dot{x}_i$ is the normalized value of the $i$th dimension in the feature vector $X$ and $x_i$ is the value before normalizing the $i$th dimension in the feature vector $X$. $x_{min}$ and $x_{max}$ are the minimum and maximum values in the feature vector, respectively.

# 4 Experiments

## 4.1 Dataset and evaluation metrics

In our experiment, we use two public datasets and a real-world dataset. Their details are as follows.

(1) ISCX 2014 [4]. This dataset is a comprehensive network traffic dataset that employs an overlay methodology to integrate three different datasets: the ISOT dataset [26], the ISCX 2012 IDS dataset [27], and a botnet traffic dataset generated by the Malware Capture Facility Project 错误!未找到引用源。. A significant advantage of this dataset is its rich variety of botnet types, which include seven C2 botnets and eight P2P botnets.

(2) CTU-13-Graphs [17]. This dataset is a graph dataset without any flow information. The dataset comprises botnet traffic extracted from the original CTU-13 [24] flow data and background traffic acquired from the 2018 CAIDA [32] dataset. The authors of the dataset balanced the numbers of legitimate and malicious nodes and divided the data into two subdatasets according to the botnet architectures: a P2P dataset and a C2 dataset. This division strategy makes the dataset suitable for training GNNs under two different architectures.

(3) Real-world dataset. This dataset was obtained from traffic collected by a national

computer network center in China. It comprises all network traffic generated from 90 suspicious IPs within China on May 12, 2023. After expert verification and threat intelligence analyses were performed, it was determined that out of the 90 IPs, 21 were legitimate, while 69 were bots. In total, this dataset consists of 1,048,575 flow records.

The compositions of the datasets used in our experiment are shown in Table 1. Graphs_C2_Ds and Graphs_P2P_Ds are used for pretraining the GCN under the corresponding architectures. The ISCX 2014 dataset is used for the final training and testing steps of the classification model. As the detection model in our study needs to separately detect botnets with C2 and P2P architectures, we must split the ISCX dataset into two subdatasets with each containing a single type of botnet architecture. Since each malicious node in the ISCX dataset is infected by only one type of bot, i.e., either a C2 or P2P bot, we can collect the P2P botnet traffic and C2 botnet traffic in different subdatasets, and we retain all legitimate traffic in the two subdatasets. In this way, we separate the ISCX dataset into C2_Ds and P2P_Ds. The real-world dataset is used to assess the performance of our method in real-world scenarios.

Table 1 The datasets used in our experiment

| Renamed Dataset | Dataset Composition | Purpose | Bot Types | #Bots |
|---|---|---|---|---|
| Graph_C2_Ds | C2 in CTU-13-Graphs[17] | Pretrain GCN for C2 | Neris, Rbot, Sogou, Virut, Menti, Murlo | 9278 |
| Graph_P2P_Ds | P2P in CTU-13-Graphs[17] | Pretrain GCN for P2P | NSIS.ay | 9620 |
| C2_Ds | C2 in ISCX [4] | C2 validation | Neris, Rbot, Virut, Menti, Sogou, Murlo, Tbot, Zeus | 30 |
| P2P_Ds | P2P in ISCX [4] | P2P validation | Zeus, NSIS, Zero Access, Weasel, Black hole, Smoke Bot, Osx_trojan, ISCX IRC bot | 16 |
| Real_World_Ds | Network traffic of 90 IPs on May 12, | Capability assessment in real | - | 69 |

We utilize various metrics to evaluate the performance of the model, including accuracy, recall, the false-positive rate (FPR), the F1 score, and the area under the receiver operating characteristic curve (ROC-AUC). Recall is the proportion of malicious nodes detected; the FPR is the proportion of legitimate nodes misclassified as bot nodes; the F1 score is the harmonic mean of precision and recall; and the ROC-AUC represents the area under the ROC curve, which intuitively describes the ability of the model to distinguish between positive and negative samples.

## 4.2 Experimental setup

The experimental codes are written in Python 3.10. The PyTorch library and PyTorch Geometric library are used for GNN training, and the sklearn library is used for machine learning model training. The experiments are executed in an Ubuntu-22.04 virtual environment equipped with a 2.50-GHz 12th Gen Intel® Core™ i5-12400 processor and 32 GB of RAM. The hyperparameters used for the GCN in the experiment are shown in Table 2.

Table 2 Hyperparameters of the GCN per training session

| No. of Layers | No. of Hidden Channels | Learning Rate | Activation Func. | Optimizer |
|---|---|---|---|---|
| 12 for C2<br>24 for P2P | 32 | 0.003 | ReLU () | Adam |

We design 5 experiments as follows.

1) **Comparison with Other Methods**: We compare our method with the state-of-the-art flow-based, topology-based and GCN-based models.

2) **Real-World Capability Assessment:** We apply our method to a real-world dataset to evaluate its effectiveness.

3) **Ablation Experiment**: We compare the detection results obtained using only flow features, purely topological features from the GCN, and the fused features of both types.

4) **Flow Feature Effectiveness Experiment**: We compare the effectiveness of the flow features used in our method to those used in other flow-based detection methods.

5) **GCN Network Experiment**: We pretrain GCNs with different numbers of layers to detect botnets with C2 and P2P architectures and identify the optimal number of GCN layers.

6) **Classifier Comparison Experiment**: We compare 5 classification models, namely, the Extra Tree, AdaBoost, naive Bayes, SVM, and MLP models, to determine which model achieves the best effect.

## 4.3 Results

**(1) Comparison with other methods**

In this experiment, we compare our method with the current best-performing botnet detection methods based on flow or topological features under the C2 and P2P architectures. We perform a 10-fold cross-validation on the C2_Ds and P2P_Ds datasets, averaging the metric results obtained from the 10 detection iterations. Under the C2 and P2P architectures, the GCN model has 12 and 24 layers, respectively, and is pretrained using the corresponding dataset.

For the comparative flow-based methods, we choose three detection models, i.e., those of Alauthaman et al. [19], D. Zhao et al. [26] and Kirubavathi et al. [11]. Additionally, we include a conversation-based detection method from Chen et al. [20]. For the comparative topology-based methods, we choose the methods of Alharbi et al. [21], which uses conventional topological features, and Zhou et al. [17], which is a GCN-based detection method. We also compare our method with that of Wang et al. [23] which is a detection model that combines flow and topological features. We briefly introduce each comparison method and its applicable botnet architectures.

**Alauthaman et al. [19]**: This method extracts a comprehensive array of flow features, such as the number of packets, average packet length, and average connection time, to train a fully connected neural network for node classification. It is designed for P2P botnet architectures.

**D. Zhao et al. [26]**: This method extracts traffic features such as the variance and mean the packet length and the packet transmission rate to train a decision tree for classification purposes. This approach is intended for C2 botnet architectures.

**Kirubavathi et al. [11]**: This method extracts only four features pertaining to small packets (with packet lengths between 40 and 320 bytes) and uses a naive Bayes classifier for classification. This method is applicable to both C2 and P2P botnet architectures.

**Chen et al. [20]**: This method extracts features such as the standard deviation of the bytes transmitted within each conversation, and the classification model is AdaBoost. This approach is suitable for both C2 and P2P botnet architectures.

**Alharbi et al. [21]**: This method extracts the in-degree, out-degree, and centrality levels of the nodes in a communication graph as topological features and subsequently uses Extra Tree to classify the nodes. This approach is suitable for C2 botnet architectures.

**Zhou et al. [17]**: This method uses only topological information acquired from the network communication graph to train a 12-layer GCN model for node classification, which can detect botnets with both C2 and P2P architectures.

**Wang et al. [23]**: This method employs two flow-based models and one graph-based model. The three models vote on malicious nodes. This method is applicable to both C2 and P2P architectures.

The comparison results are shown in Table 3 (for C2 architectures) and

Table 4 (for P2P architectures). In both types of architectures, the node classification model using a GCN [17] is invalid because of the dataset imbalance problem. Thus, the F1 score of this method cannot be calculated. This also shows the necessity of pretraining the GCN on a balanced dataset. Among the other comparison methods, for C2 architectures, the detection effects of methods introducing topological features are generally better than those of flow-based methods. Our method achieves the best results in terms of all the metrics. With respect to the P2P architectures, our method achieves the highest accuracy and recall, and its F1 score is the second-highest value. Overall, our method has a better ability to identify malicious nodes while ensuring high precision.

In terms of the time cost, both flow-based and conversation-based methods have time overheads below *1 s*, while the method based on traditional graph features has a time overhead of nearly *50 s*, which is significantly longer than those of the other methods. Our method and the method using a GCN have time overheads of *1~2 s*, satisfying the requirements of real-time detection.

Table 3 Comparison with other methods based on the C2 topology

| Method | Type | Accuracy | Recall | F1 | Running time (*s*) |
|---|---|---|---|---|---|
| D. Zhao et al. [26] | Flow | 0.8318 | 0.8251 | 0.8274 | 0.492 s |
| Chen et al. [20] | Conversation | 0.8884 | 0.7810 | 0.8226 | 0.627 s |
| Kirubavathi et al. [11] | Flow | 0.8065 | 0.7651 | 0.7829 | **0.403 s** |
| Alharbi et al. [21] | Topology | 0.9779 | 0.8176 | 0.8534 | 46.333 s |
| Zhou et al. [17] | Topology (GCN) | 0.9931 | 0.0 | - | 1.710 s |
| Wang et al. [23] | Topology + Flow | 0.9698 | 0.8823 | 0.8333 | 21.230 s |
| **Ours** | Topology + Flow | **0.9885** | **0.9290** | **0.9276** | 1.997 s |

Table 4 Comparison with other methods based on the P2P topology

| `Method | Type | Accuracy | Recall | F1 | Running time (*s*) |
|---|---|---|---|---|---|
| Alauthaman et al. [19] | Flow | 0.9730 | 0.9126 | **0.9419** | **0.145 s** |
| Kirubavathi et al. [11] | Flow | 0.9470 | 0.8337 | 0.8755 | 0.232 s |
| Chen et al. [20] | Conversation | 0.8928 | 0.8733 | 0.8734 | 0.307 s |
| Zhou et al. [17] | Topology (GCN) | 0.9930 | 0.0 | - | 1.513 s |
| Wang et al. [23] | Topology + Flow | 0.9713 | 0.5384 | 0.5833 | 18.230 s |
| **Ours** | Topology + Flow | **0.9910** | **0.9466** | 0.9235 | 1.801 s |

**(2) Assessment of real-world capabilities**

The architectures of the botnets in Real_World_Ds are C2 architectures; thus, the GCN trained on the corresponding botnets is used for feature fusion. Tenfold cross-validation is employed to assess the performance of the model. The results are shown in Table 5.

Table 5 Model assessment results obtained on a real-world dataset

| Accuracy | Recall | F1 | Running time ($s$) |
|---|---|---|---|
| 0.8341 | 0.8496 | 0.8477 | 5.770 s |

The real-world dataset incorporates normal daily network traffic and real botnet communications, posing additional challenges for detection. Therefore, although the model performance slightly declines compared to that achieved in experiment (1), this experiment still demonstrates the effectiveness of the model in real-world scenarios.

**(3) Ablation experiments**

We compare the detection effects of models using (1) only flow features, (2) only topological GCN features, and (3) the fusion of both kinds of features. The C2_Ds dataset is used for the C2 architecture, and the P2P_Ds dataset is used for the P2P architecture. The detection results are validated using a 10-fold cross-validation method. The experimental results shown in Figure 3 and Figure 4 demonstrate that the model using the fused features is significantly superior to the model using only topological features based on both architectures. On the other hand, compared to using only the flow features, utilizing the fused features significantly improves the recall and F1 score attained under both architectures, indicating that the latter model has a better ability to identify malicious nodes. These results prove the effectiveness of the feature fusion method proposed in this paper.

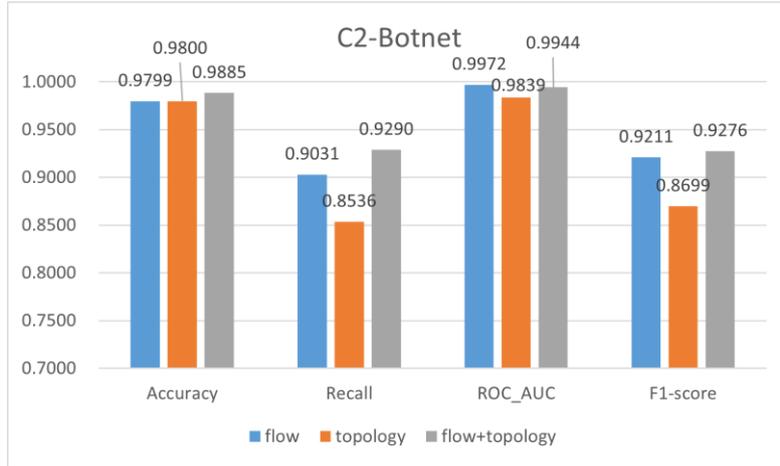

Figure 3 Ablation experiment conducted under the C2 topology.

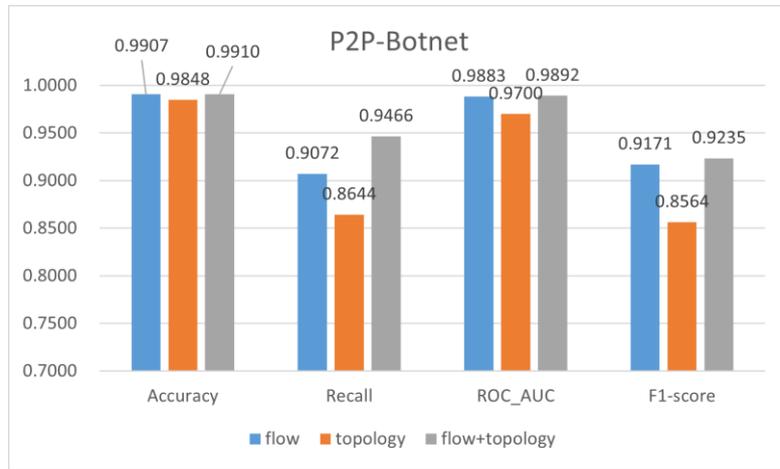

Figure 4 Ablation experiment conducted under the P2P topology.

**(4) Flow feature effectiveness experiment**

In this experiment, we aim to illustrate the effectiveness of the five selected flow features. We employ a 10-fold cross-validation method on C2_Ds and P2P_Ds to compare the flow features used herein with those used by other flow-based methods. Figure 5 shows the specific comparative results. With respect to the C2 botnet architecture, the features designed in this paper significantly outperform those of the other methods in terms of all the metrics. Concerning the P2P botnet architecture, the features designed here also achieve excellent results that are almost identical to those of the best-performing features designed by Alauthaman et al.[19]. The time overhead comparison among the tested models when extracting each feature in a one-minute time window is shown in Table 6. The five features designed here require the least time consumption. Overall, the five features used in our method have the shortest processing time and exhibit good detection performance for botnets with different architectures.

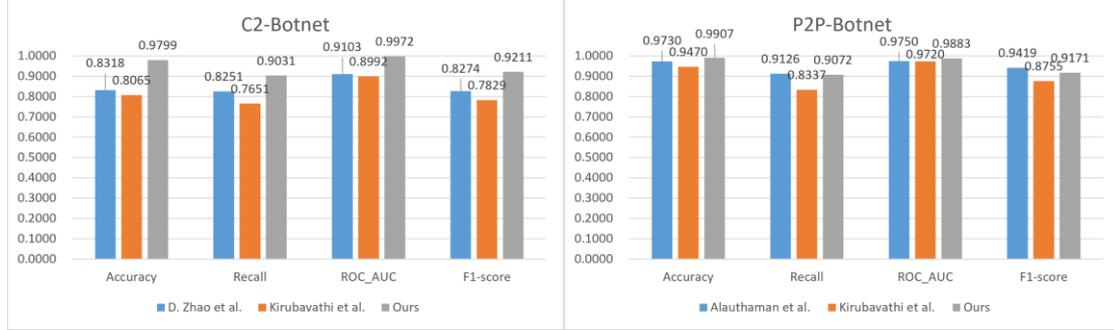

Figure 5 Flow feature effectiveness comparison under the C2 and P2P architectures.

Table 6 Average feature extraction times required by the proposed method and other methods

| Method | Running time (s) |
| --- | --- |
| Alauthaman et al. [19] | 0.145 |
| Kirubavathi et al. [11] | 0.299 |
| D. Zhao et al. [26] | 0.503 |
| **ours** | **0.088** |

**(5) GCN experiment**

This experiment aims to determine the optimal number of GCN layers for different network architectures. Specifically, we pretrain GCNs with varying numbers of layers on Graph_C2_Ds and Graph_P2P_Ds. After the pretraining process, we employ the GCN models to extract fusion features from the corresponding architectural datasets (C2_Ds/P2P_Ds). Finally, we train an Extra Tree model to evaluate the effects of different fusion features derived from GCNs with various layer counts.

For the C2 architecture, we experiment with 10, 12, 14, and 16 GCN layers. Given the more divergent structure of the P2P architecture, which requires a deeper network to extract topological features, we test GCNs with 16, 20, 24, and 28 layers.

Figure 6 and Figure 7 show that under the C2 architecture, the most suitable number of GCN layers for feature fusion is 12; under the P2P architecture, the malicious node detection rates are roughly equivalent when the number of GCN layers varies.

However, fusion the features acquired from a 24-layer GCN results in the lowest FPR and the highest AUC-ROC. In summary, we choose a 12-layer GCN for the C2 architecture and a 24-layer GCN for the P2P architecture.

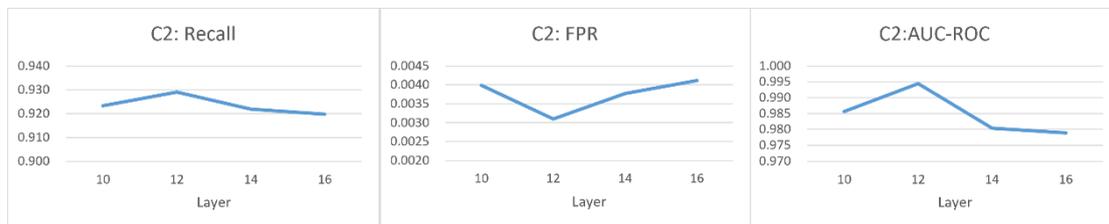

Figure 6. Under the C2 architecture, detection effectiveness achieved by fusing features from various layers of the GCN

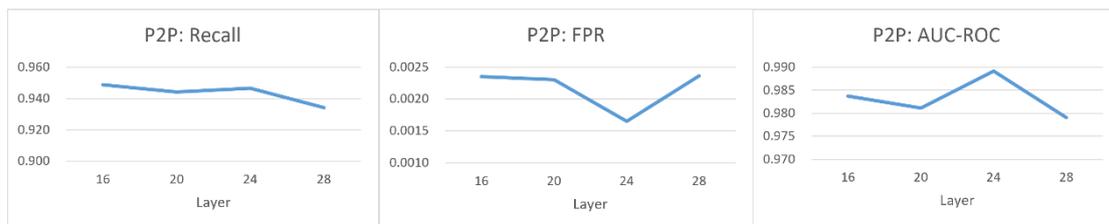

Figure 7 Detection effectiveness achieved under the P2P architecture using features fused by GCNs with various numbers of layers..

**(6) Classifier comparison experiment**

In this experiment, we select five widely used and effective machine learning models, i.e., the Extra Tree, AdaBoost, naive Bayes, SVM, and MLP models, with 12 hidden layers. We test the classification performance of the different models on C2_Ds using the fused features. The results in Table 7 show that the Extra Tree classifier performs best overall, so we choose it as our final classifier.

Table 7 Comparison among the detection effects of different machine learning models

| Classifier | Recall | FPR | AUC |
| --- | --- | --- | --- |
| **Extra Tree** | **94.66%** | **0.16%** | **98.92%** |
| AdaBoost | 94.47% | 3.60% | 98.55% |
| Naive Bayes | 83.42% | 10.85% | 94.85% |
| SVM | 66.17% | 21.59% | 67.59% |
| MLP (12 hidden layers) | 82.86% | 0.17% | 97.89% |

# 5 Conclusions

In this paper, we propose an innovative framework that contains a pretrained GCN model and an Extra Tree classification model for botnet detection. We leverage the message propagation mechanism of the GCN to extract topological features and fuse them with flow features for detection purposes. Our method can be used for botnet detection under C2 and P2P architectures; only the number of GCN layers must be adjusted.

First, we design five easily accessed yet hard-to-evade flow features. Then, we construct network communication graphs that integrate the flow features and the connection topologies of nodes. To address the issue of a GCN training on imbalanced datasets, we pretrain the GCN using balanced graph datasets, enabling its capacity for topological feature extraction. This pretrained GCN is subsequently used for feature fusion. By using varying numbers of GCN layers, our model demonstrates flexibility in terms of detecting botnets with both C2 and P2P architectures. Experiments conducted on public datasets indicate that our method outperforms the current state-of-the-art methods under C2 and P2P architectures. We also apply our model to a real-world detection task and achieve satisfactory results.

In future work, we aim to expand our model to enable the simultaneous detection of botnets with both C2 and P2P architectures, thereby extending its applicability.